\documentclass[twocolumn,footinbib,showpacs]{revtex4-1}
\usepackage{graphicx}
\usepackage{amsmath}
\begin{document}
\title{Econophysics of adaptive power markets: \\ When a market does not dampen fluctuations but amplifies them}Ê
\author{Sebastian M.\ Krause}
\author{Stefan B\"orries}
\author{Stefan Bornholdt}
\affiliation{Institut f\"ur Theoretische Physik, Universit\"at Bremen, D-28359 Bremen, Germany}
\begin{abstract}
The average economic agent is often used to model the 
dynamics of simple markets, based on the assumption that the 
dynamics of many agents can be averaged over in time and space. 
A popular idea that is based on this seemingly intuitive notion 
is to dampen electric power fluctuations from fluctuating  
sources (as e.g.\ wind or solar) via a market mechanism, 
namely by variable power prices that adapt demand to supply.  
The standard model of an average economic agent predicts 
that fluctuations are reduced by such an adaptive pricing mechanism. 

However, the underlying assumption that the actions of all 
agents average out on the time axis is not always true
in a market of many agents.  
We numerically study an econophysics agent model of an adaptive 
power market that does not assume averaging a priori.  
We find that when agents are exposed to source noise via 
correlated price fluctuations (as adaptive pricing schemes suggest), 
the market may amplify those fluctuations. In particular, 
small price changes may translate to large load fluctuations 
through catastrophic consumer synchronization. 
As a result, an adaptive power market may cause the opposite 
effect than intended: Power fluctuations are not dampened 
but amplified instead. 
\end{abstract}
\maketitle

\section{Introduction} 
Modern power markets face the challenge to satisfy a continuous demand 
for electricity, despite fluctuating energy sources as, e.g.\ solar or wind 
\cite{pepermanns2005,carrasco2006,blaabjerg2006,kundur1994,butler2007,marris2008}. 
It has been proposed to reduce fluctuations in power markets 
via time-varying pricing schemes, in order to stimulate the shift of 
energy consuming activities with flexible execution times as, e.g. 
washing or heating, to times with excess supply 
\cite{luh1982,allcott2011,agarwal2012,roozbehani2012}.
From the perspective of a standard economic theory this is a simple 
picture: A specific value of the price leads to a predictable total 
demand. Consequently, there is an equilibrium price, 
where demand and supply are balanced. As a result, one would expect 
that part of the demand thereby is shifted to times with lower prices
\cite{luh1982}. 
Thus the market would act as a low pass filter for power fluctuations,
an elegant idea at first sight, indeed. 

However, real markets often behave differently than the single 
representative agent of standard economic theory, most prominently 
illustrated by crashes of stock markets and similar phenomena 
resulting from interactions among many agents 
\cite{kirman1989,kirman1993,challet1997,lux1999,sornette2004}. 
Even in markets where agents do not interact directly, they may exhibit 
coordinated behavior. 
For example, the actions of consumers may self-organize 
on the time axis, with catastrophic synchronization as a possible result. 
In that case, averaging over the dynamics of many agents over time  
is not appropriate because the central limit theorem does 
not hold anymore. The market, instead of acting like a low pass filter 
that dampens fluctuations, turns into a generator for catastrophic time series. 

In fact, problems with the central limit theorem in dynamical systems 
with many degrees of freedom are well known from different fields, 
and often are related to time series that exhibit large fluctuations.  
Such phenomena have been discussed, for example, in the contexts of 
earthquakes, rice piles, stock markets, solar flares, and mass extinctions 
\cite{newman1996,newman1996_2,sneppen1997}.  
These systems have in common that fluctuations with broad or power law 
size distributions occur that do not need a full mechanism of 
self-organized criticality (SOC) at work. Instead, coherent stochastic 
noise acting on a system with many agents may suffice to explain 
such power law distributed fluctuations \cite{newman1996}. 
Agents can react to the coherent noise in a way that causes their 
actions to synchronize at rare events. As a result, power law distributed 
event sizes appear even for narrow (and even Gaussian) distributions 
of the coherent noise \cite{sneppen1997}. 

In this paper we study whether this mechanism may be at work 
in markets, or more specifically, in power markets.
Collective behavior of agents in a market can be treated with agent 
based models  allowing for individual behavior of agents.
Agent based models constructed on simple rules of individual behavior 
in markets have been shown to exhibit many features of real markets 
\cite{kirman1993,challet1997,lux1999,sornette2004,samanidou2007}. 
We here study one of the simplest possible agent based models for 
an adaptive power market. 

Our toy model consists of independent agents reacting to a predefined 
global price time series. Their rare consumption events set in, 
once the actual price is below an individual highest acceptable price. 
The highest acceptable prices of each agent are updated with a stochastic 
process to account for saturation after consumption and growing need for 
electricity in times without consumption. 
This is to model rare consumption events with flexible execution time, 
while the base demand connected to time-fixed activities is ignored 
in this study. 
We analyze the effect of demand synchronization at low prices. 
As a result, the total demand can exceed the average demand by several orders of magnitude. 
To prove the robustness of this behavior, we analyze the demand distribution 
and the demand curve (demand over price) for different price time series 
with and without correlation. We find the behavior of our artificial market 
to be in sharp contrast to standard economic theory. 
A sensitive demand curve and saturation effects question the 
application of equilibrium prices. 

\section{\label{sec:model}Model description}
\begin{figure}[htb]
\begin{center}
	\includegraphics[width=0.85\columnwidth]{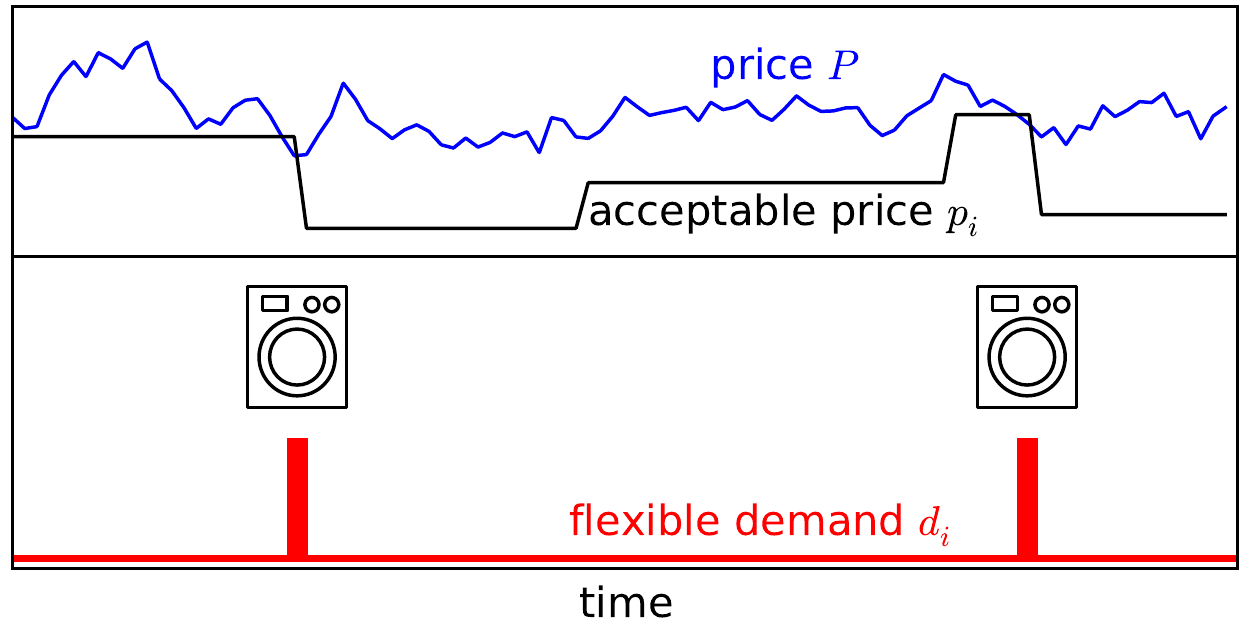}
	\caption{Sketch of the model dynamics for an indiviual agent. 
	Top: Price time series (blue) together with the price acceptance 
	of the agent (black). Bottom: Demand of the agent. 
	With increasing time we see a consumption event with 
	lowering the price acceptance (saturation), followed 
	by two increments (growing need) and another consumption.}
	\label{fig:model}
\end{center}
\end{figure}
Let us now define the power market agent model. 
We analyze an artificial market consisting of one power provider and $N$ power consumers. 
In every (integer) time step $t$ the power provider sets a price $P(t)$ 
(with time average $\bar P=1$) visible to all consumers. 
The individual demand of an agent $i$ is defined as 
\begin{align}
d_i(t) &= \begin{cases}1 & P(t)\leq p_i(t)\\ 0 & P(t)>p_i(t)\end{cases}
\end{align}
with its individual highest acceptable price $p_i(t)$, 
as illustrated in Fig.~\ref{fig:model}. 
It is initialized with $p_i(t=0)={\rm rand}(0,1)$ and evolves according to 
\begin{align}
p_i(t+1) &= \begin{cases}
       {\rm rand}[0,p_i(t)], & 
       \textrm{if }P(t)\leq p_i(t), \\
       {\rm rand}[p_i(t),1], & 
       \textrm{else with prob. } f,\\
       p_i(t), & {\rm else}.\end{cases}
\end{align}
The term ${\rm rand}(a,b)$ denotes a random number uniformly drawn 
from the half open interval $[a,b)$. The first case ($P(t)\leq p_i(t)$) 
corresponds to power consumption at time $t$. As a consequence, 
the acceptable price will then also be lowered to represent saturation. 
The second case, rare increases of the highest acceptable price 
$p_i$ with probability $f\ll 1$, is to model the increasing need 
for power-consumption with time. 
This stochastic evolution of $p_i(t)$ is inspired by the coherent 
noise model by Newman and Sneppen \cite{newman1996} where a 
resilience threshold towards catastrophic events is evolved in time. 

The total demand $D(t)=\sum_{i=1}^N d_i(t)$ is satisfied by the power provider. 
We avoid including an additional contribution of time-fixed activities 
$D_{\rm base}(t)$ into this model, since this part is not the focus 
of the present study and would not change the overall dynamics. 
To analyze the capabilities of the power provider 
to shape demand time series $D(t)$, we use different types of noisy time series $P(t)$. 
We take independent identically distributed prices out of a Gaussian 
distribution with mean $\bar P=1$ and different standard deviations $\sigma_P$. 
Additionally, to consider correlations over time (as they are known for 
common price time series and for weather phenomenons), 
we use a Langevin-equation 
\begin{align}
P(t+1) - P(t) &= -v_0 \cdot (P(t)-\bar P)+\sigma_0 \cdot \xi(t)\label{eq:langevin}
\end{align}
with an independent normally distributed random variable $\xi(t)$ 
(the green line in Fig.~\ref{fig:dichten} shows the Gaussian density for such a time series). 

\section{\label{sec:sync}Synchronization at low prices}
\begin{figure}[htb]
\begin{center}
\includegraphics[width=1.0\columnwidth]{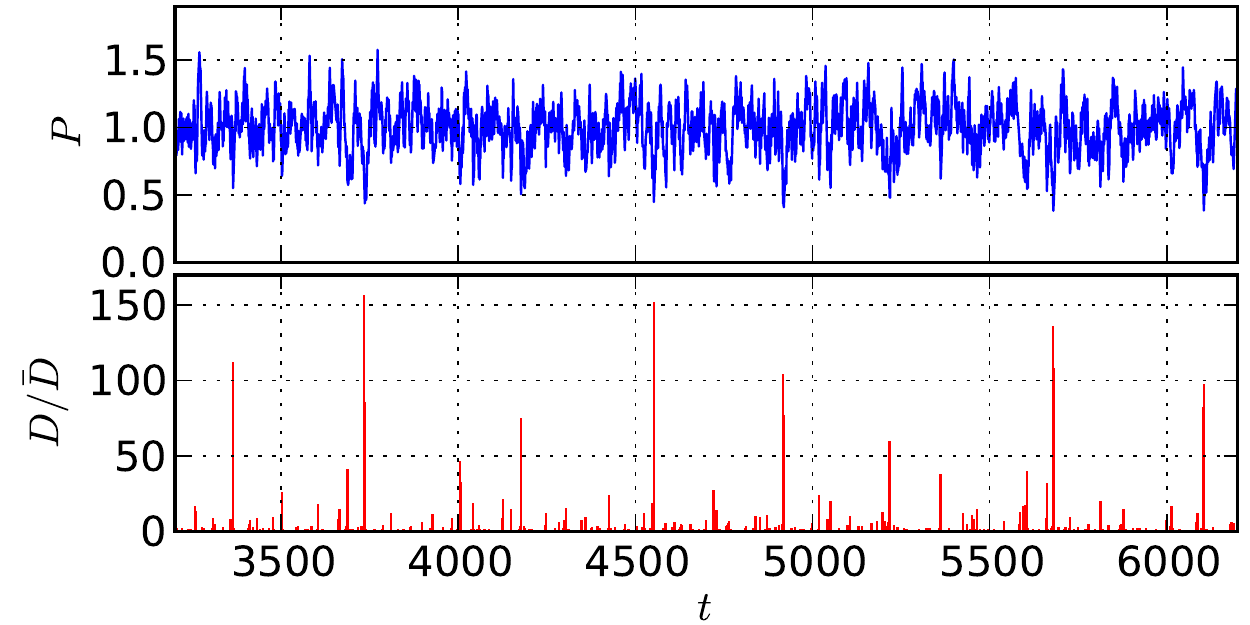}
\caption{Total demand compared to the time-averaged total demand 
$D/\bar D$ (bottom) of the system reacting to a correlated time series (top). 
At low prices, consumers execute their rare consumption activities 
in a synchronized fashion, leading to total demands $D$ far above 
the average demand.}
	\label{fig:zeitreihe}
\end{center}
\end{figure}
Fig.~\ref{fig:zeitreihe} on top shows a section of a price time series 
according to Eq.~(\ref{eq:langevin}) with $v_0=0.2$ and $\sigma_0=0.1$. 
In the bottom panel, we see the according demand divided by the average demand. 
The average demand $\bar D=\frac{1}{T+1}\sum_{t=1}^{T}D(t)$ for the system 
with $f=10^{-3}$, $N=10^6$ agents and a simulation time of $T=10^7$ 
(plus $10^3$ initial time steps for reaching a stationary state) was 
calculated to be $\bar D=979$. Therefore, a single agent demands 
on average $\bar d=\bar D/N =9.79\cdot 10^{-4} \approx f$. 
The time series of $D(t)/\bar D$ shows demand peaks more than two orders 
of magnitude above the average demand dominating the whole time series. 
This is due to synchronization: At low prices, many agents 
demand at the same time. 
As a result, the prices, fluctuating in a narrow range, cause a broadly 
distributed demand time series with extreme events. 

\begin{figure}[htb]
\begin{center}
\includegraphics[width=1.0\columnwidth]{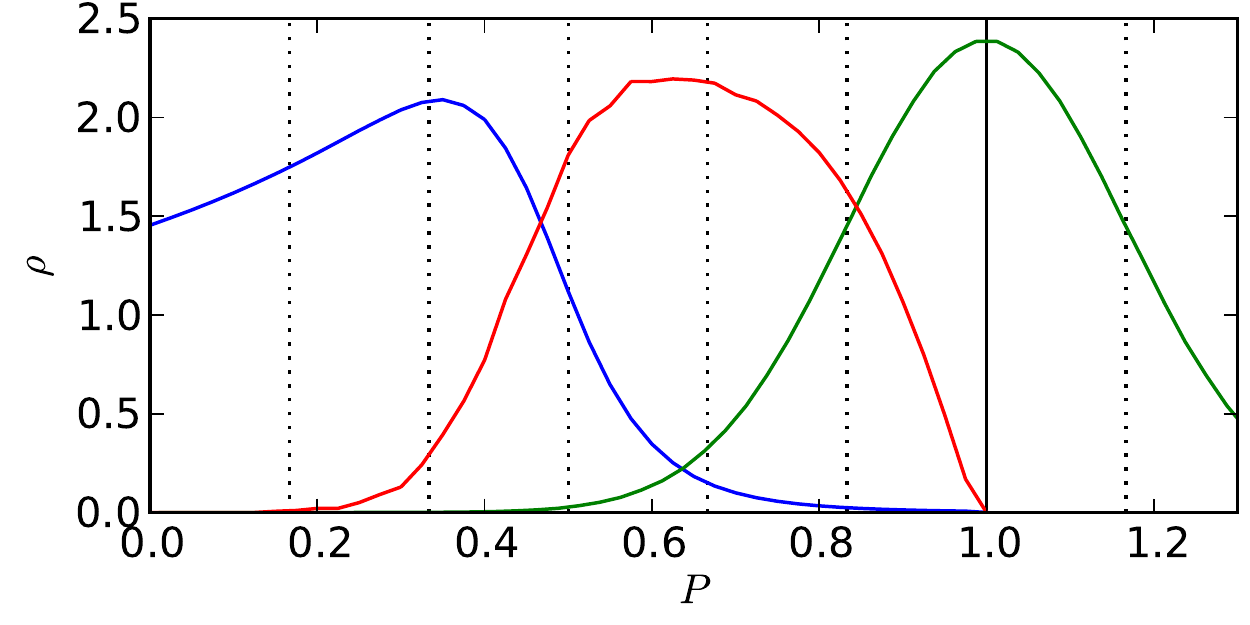}
\caption{Densities of the highest acceptable prices $p_i(t)$ 
(blue line), of the total load consumed at certain prices 
(red line) and of the prices (green line).}
\label{fig:dichten}
\end{center}
\end{figure}
In Fig.~\ref{fig:dichten} we see the density of highest acceptable 
prices $p_i(t)$ averaged over time and agents (blue line), 
\begin{align}
	\rho_p(P) & =\sum_{i=1}^N\sum_{t=0}^T \Delta_{p_i(t),P}/N(T+1), 
\end{align}
the density of total loads consumed at certain prices (red line),
\begin{align}
	\rho_D(P) &= \sum_{t=0}^T D(t)\Delta_{P(t),P}/\bar D,  
\end{align}
and the price distribution (green line). 
With $\Delta_{p,P}=1$ for $p$ and $P$ lying in the same interval 
and $\Delta_{p,P}=0$ else, a binning of values is realized. 
The average price $\bar P$ is indicated with a black vertical line, 
and multiples of one standard deviation of the price distribution 
are indicated with dotted vertical lines. 
We observe that only a small fraction of the demands are executed 
within one standard deviation of the price, 
35\% of the price events only lead to 18\% of the demand. 
This part is due to agents who need to consume power very soon. 
The average price for consumers $\sum_t P(t)D(t)/\sum_t D(t) = 0.65$ 
is much lower. Due to synchronization effects, rare events below 
$\bar P - 3\sigma_P=0.5$ constituting only 0.14\% of the time 
series lead to a part of 16\% of the total demand. 
In conclusion the agents indeed consume at low costs and 
their strategy is beneficial. Additionally, the strategy represents 
individual needs, implemented by random moves of the individual 
highest acceptable prices. 

\section{\label{sec:provider}Robust occurrence of high demand}
\begin{figure}[htb]
\begin{center}
\includegraphics[width=1.0\columnwidth]{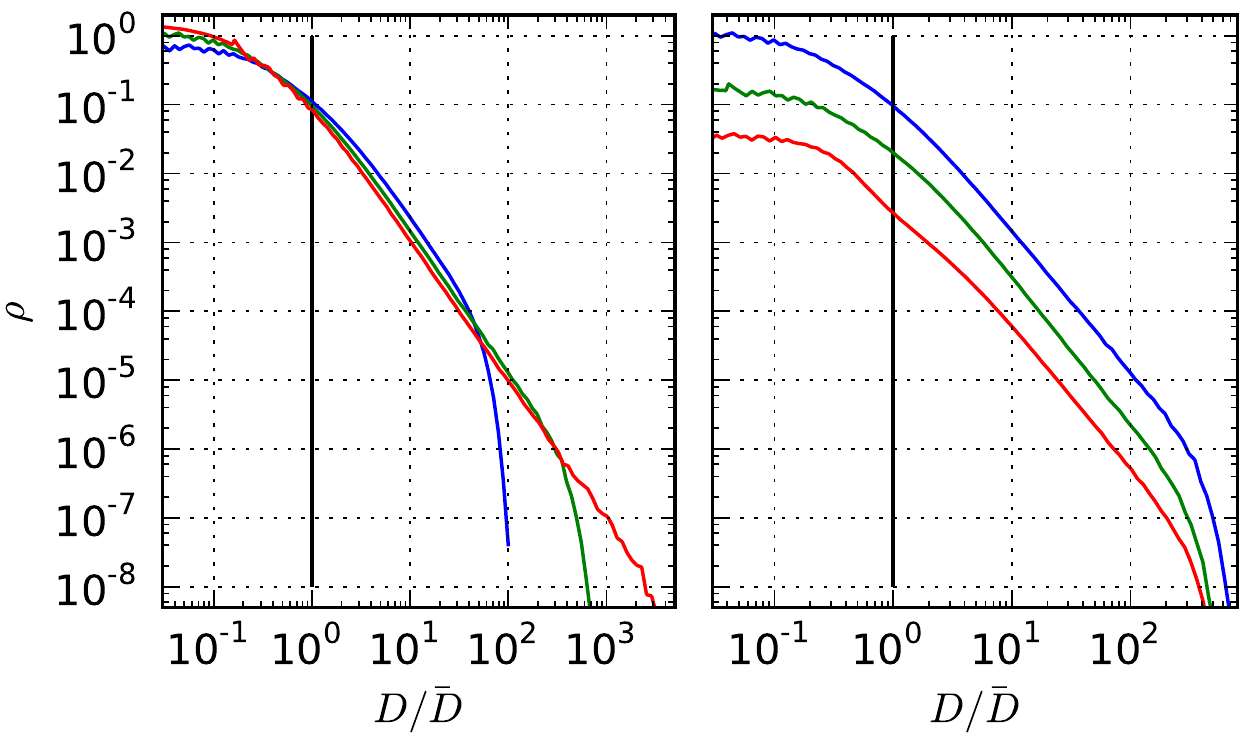}
\caption{Demand density $\rho(D/\bar D)$ for different parameters 
$f$ ($f=10^{-2}$, $f=10^{-3}$ and $f=10^{-4}$, left) 
and for different price time series (right). 
All distributions are broad with frequently occuring events of large demand.}
\label{fig:lastverteilung}
\end{center}
\end{figure}
In Fig.~\ref{fig:lastverteilung} on the left, the distribution 
of demand $D$ is shown for independent Gaussian distributed prices 
with $\sigma_P=1/6$ and different scarcity of consumption 
($f=10^{-2}$, $f=10^{-3}$ and $f=10^{-4}$). 
All simulations in this study are done with $N=10^6$ and $T=10^7$. 
Even in the case $f=10^{-2}$, where consumers buy on average 
in one of hundred time steps, maximum demands are almost 
two orders of magnitude larger than the average demand. 
For rarer consumption (smaller values of $f$), 
the distribution of loads clearly gets the shape of 
a truncated power law with increasing cutoff for $f\to 0$, 
as expected from \cite{newman1996}. 

The results for different price time series shown on the right 
of the figure emphasize the robustness of synchronization 
in our artificial market. 
The results are shifted for better visibility. 
On top the result for Gaussian distributed prices with 
$\sigma_P=1/6$ is shown again. 
Below  the same type of price time series with $\sigma_P=1/20$ 
is used with very similar results. Changing the standard 
deviation of the prices $\sigma_P$ leads to the same dynamics, 
only with buying events at different typical prices. 
It is known from \cite{sneppen1997}, that using other 
distributions for the prices does not change the results considerably. 
The graph below shows the result for the correlated 
time series of Sec.~\ref{sec:sync} (Eq.~(\ref{eq:langevin}) 
with $v_0=0.2$ and $\sigma_0=0.1$). 
The same type of broadly distributed demand emerges. 
We also tested a real price-like time series by using 
daily closure values of the Dow Jones (1900-2007, detrended data) 
with similar results (not shown). In conclusion this means 
that for our artificial market the synchronization of 
consumers occurs for very different price time series and can hardly be avoided. 

\begin{figure}[htb]
\begin{center}
\includegraphics[width=1.0\columnwidth]{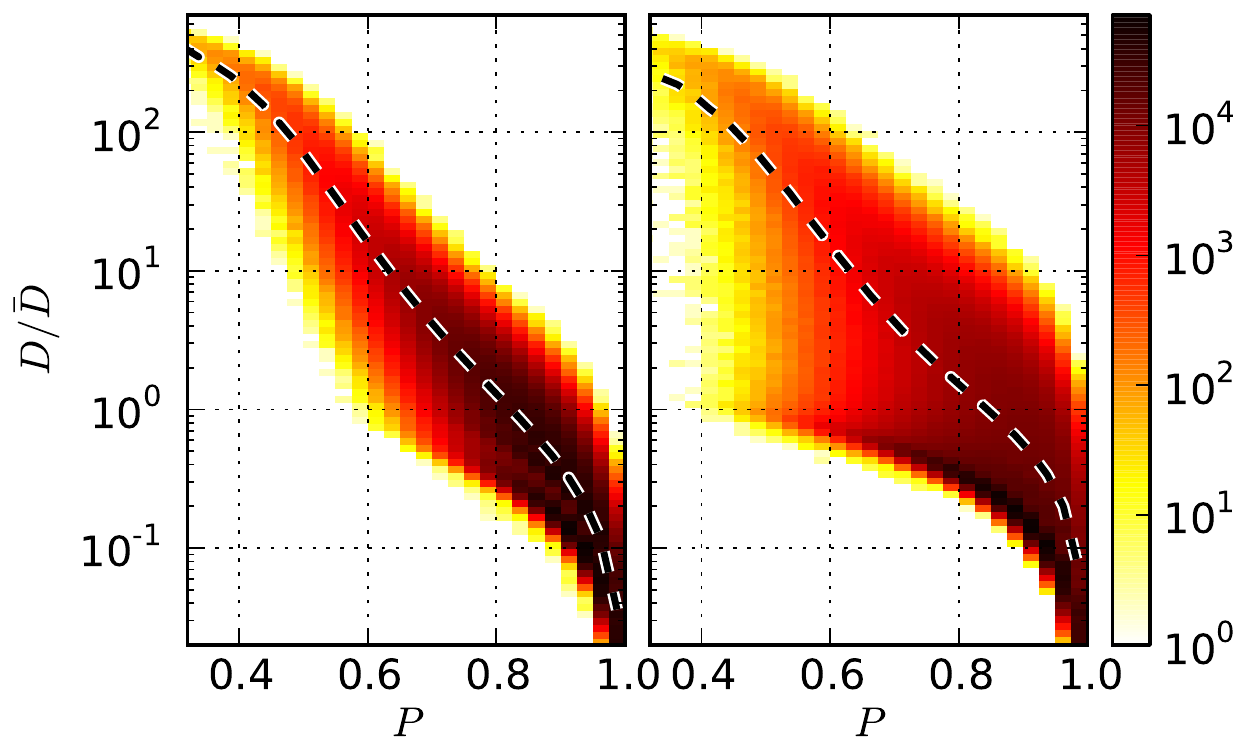}
\caption{Binning of events by price intervals and demand intervals. 
Relative frequency of events is shown with color values in 
logarithmic scale for Gaussian distributed uncorrelated prices 
(left) and correlated prices (right). The dashed lines show 
the average demand for price intervals (the so-called demand 
curve) with an exponential dependence in both cases.}
	\label{fig:absatzkurve}
\end{center}
\end{figure}
Finally let us discuss the demand curve of the power provider. 
In Fig.~\ref{fig:absatzkurve} we see a binning of events 
according to the rescaled demand $D/\bar D$ and price $P$. 
The counts of events are shown with color values in logarithmic 
scale. On the left we see the results for uncorrelated 
prices and $\sigma_P=1/6$. Due to the distribution of 
prices $P$, events with low prices are generally rare, 
but if they occur, they lead to high demand. 

The average demand for a certain price interval according 
to this binning is indicated with a dashed line. This is 
the so-called demand curve frequently used in standard 
economics to calculate equilibrium prices. 
The demand spans more than three orders of magnitude 
within about four standard deviations of the price 
(we checked that the same holds for the smaller 
value $\sigma_P=1/20$). Smooth changes of the price 
lead to drastic changes of the demand. This is in 
sharp contrast to standard economics and limits the 
feasibility of equilibrium prices. 
Additionally, the demand values span more than an 
order of magnitude for many price values. 
This is due to saturation effects (only few 
agents buy at a low price, if a lower price 
recently occurred). 

In the right panel of Fig.~\ref{fig:absatzkurve}, 
results for the correlated price time series are shown. 
The demand curve is not changed considerably, 
while the distribution of demands for certain 
prices is broadened. Due to saturation effects, 
consecutive low prices lead to shrinking loads. 
 
\section{Summary and Outlook}
We studied a simple agent based model of an electricity 
market with variable prices and studied collective 
effects when consumers aim for lowest prices. 
In particular we consider consumption with 
time-flexible execution as, e.g. washing or heating. 
Time-variable consumption is modeled with a stochastic 
process for individual highest acceptable prices. 
As a central quantity, the total demand emerging in 
our artificial market has been analyzed. 

Our main observation is that the rare consumption events 
of the consumers in the market tend to strongly 
synchronize at low prices. This leads to peak demands 
exceeding the average demand by several orders of magnitude. 
These frequent extreme events account for a considerable 
part of the average demand over time. We find that high 
demands occur robustly for different types of price time series, 
as long as the pricing noise hits the consumers coherently. 
We find power law distributed demands with large extreme events,  
both, for uncorrelated price time series as well as for correlated 
time series. The catastrophic behavior of the system appears to be 
hardly to prevent. 

Finally we question the concept of equilibrium prices 
in the context of our artificial market. As the system shows 
an exponential growth of demand when prices drop, equilibrium 
prices can hardly establish. 
Demands take on a wide range of values, even at the same price. 

While these are results from a statistical physics inspired toy 
model for an electricity power market with fluctuating energy sources 
and an adaptive pricing scheme, they may provide a lesson for real 
markets as well. In particular, they seem to indicate that the, at first sight, 
brilliant idea to use market mechanisms as a low pass filter for 
fluctuating electricity sources may not only break down under 
certain conditions. More importantly they also can lead to catastrophic 
consequences when a basic prerequisite fails: Breakdown of 
the central limit theorem when consumers do not act statistically 
independently.

\end{document}